# Search for Pulsed Emission in Archival VERITAS Data

**Avery Archer*  for The VERITAS Collaboration[†]**
*Washington University in St. Louis*
*E-mail:* a.archer@wustl.edu

Since the 2011 VERITAS discovery of very high energy (VHE; E>100 GeV) gamma rays from the Crab pulsar, there has been concerted effort by the gamma-ray astrophysics community to detect other pulsars in the VHE band in order to place better constraints on emission models. Pulsar modelling demonstrates that much of the magnetosphere is opaque to VHE photons, limiting emission regions to the outer magnetosphere or beyond the light cylinder. The locations of 19 known pulsars have been observed by VERITAS since full observations began in 2007 with 11 locations having more than 20 hours of observations. Observations of VHE emission from more sources could provide key data to help constrain current models of emission location and mechanisms. We present the status of the ongoing VERITAS program searching for pulsed emission in archival data.

---

*Speaker
[†] http://veritas.sao.arizona.edu

**1. Introduction**

In 2011 the VERITAS and MAGIC collaborations announced the first discovery of pulsed emission from a pulsar, the Crab pulsar, in the VHE regime [1], [2]. Prior to these observations, 46 gamma ray pulsars were known to exist with 7 having been detected by instruments on board the CGRO [3],[4] and the remaining 39 newly discovered by the *Fermi*-LAT [5][‡]. The spectra of these pulsars were thought to have an exponential cutoff bolstering the curvature radiation scenario which predicts a spectral cut-off of this shape [6]. The discovery of VHE gamma ray pulsations from the Crab pulsar has challenged current emission mechanism models. The curvature radiation scenario alone does not adequately explain the VHE emission. A combined spectral energy distribution (SED) from *Fermi*-LAT and VERITAS, over the energy range 100 MeV to several hundred GeV, clearly favors a broken power law of the form $A \times (E/E_0)^\alpha / [1 + (E/E_0)^{\alpha-\beta}]$ [1]. Many alternative emission scenarios have been proposed to explain the VHE emission from the Crab pulsar, including inverse-Compton (IC) scenarios in the outer magnetosphere [7,8] or beyond the light cylinder [9]. The discovery of, or strong observational upper limits on, VHE emission from other pulsars will provide critical measurements of the shape of the emission spectrum above the apparent GeV break and thus strongly constrain plausible emission scenarios.

Since the discovery of pulsed emission from the Crab pulsar, there has been a concerted and ongoing effort by the gamma-ray astronomy community to add to the VHE pulsar catalogue and constrain emission models. The Vela pulsar has been detected by HESS [10] above 30 GeV and by the *Fermi*-LAT above 50 GeV [11]. VERITAS has recently reported observations of the Geminga pulsar, the second-brightest steady, high-energy (E > 100 MeV) gamma-ray source in the sky. These observations yield no evidence for periodicity in the VHE band. Upper limits above 166 GeV are reported here [12]. The Crab pulsar remains the only currently known VHE emitting pulsar. A recent stacked analysis of 115 *Fermi*-LAT pulsars found no significant cumulative excess above 50 GeV. The stacked analysis constrains the average flux of the 115 pulsars analyzed but does not exclude the possibility of finding pulsars which are as bright as the Crab pulsar above 50 GeV [13].

The locations of 19 *Fermi*-LAT pulsars have been observed since full observations with VERITAS began in 2007. Each of these pulsars is a new candidate VHE source and most have not been observed by other VHE instruments. Many of these observations were obtained when the pulsar was in the field of view of another primary target. 11 of the locations have more than 20 hours of observations each. VERITAS has the capability of detecting a 1% Crab Nebula flux from a point source in roughly 25 hours. The top 10 pulsars, in terms of spin-down power divided by distance squared, detected in the Northern Hemisphere by *Fermi*-LAT are contained in this archival dataset. In this paper we present the ongoing effort by VERITAS to search for pulsed emission from these known pulsars and add to the VHE pulsar catalog.

**2. *Fermi*-LAT Analysis**

A detailed *Fermi*-LAT analysis has been performed on each pulsar listed in Table 1 using 5.2 years of *Fermi*-LAT data to produce phase-resolved SEDs[§]. The phase-resolved SEDs include relevant phase cuts in order to avoid contamination from un-pulsed emission (e.g. pulsar wind nebula). These SEDs (not included in this proceeding) provide a measurement of the spectral shape above the break energy to enable possible extrapolations of flux in the VHE band.

---

[‡] In the recent 3rd *Fermi*-LAT Source Catalog (3FGL) the number of known gamma ray pulsars has increased to 143 [14].

[§] The size of the data set used to produce the SEDs is determined by the period of validty of the pulsar timing solutions used.

Presented in Figure 1. are the phase-averaged SEDs for 18 of these 19 pulsars as measured in the Fermi Large Area Telescope Third Source Catalog (3FGL) [15]. The procedure for creating the SEDs is described in the 3FGL. These SEDs have no phase-cut applied to exclude steady-state emission, thus the spectral shapes may shown may be a combination of pulsed and un-pulsed nebula emission.

For most pulsars, the spectral shape above the break is sufficiently well measured to preclude extrapolations which connect smoothly to the *Fermi*-LAT SED but predict a VHE flux above 1% of the Crab Nebula flux. Given this, VERITAS will adopt a broad search strategy, which includes investigations of VHE emission which manifests as a new spectral component, not connected to the *Fermi*-LAT SED. Such a ccomponent may arise due IC scattering. [16].

| Source | Exposure (min) | Average Elevation (deg) | Average Offset (deg) | Spin-down Power, $\dot{E}$ ($10^{35}$ erg s$^{-1}$) |
|---|---|---|---|---|
| PSR J0007+7303 | 2125 | 47 | 0.71 | 4.5 |
| PSR J0023+0923 | 924 | 63 | 0.55 | 0.2 |
| PSR J0205+6449 | 2043 | 55 | 0.46 | 264.4 |
| PSR J0218+4232 | 1706 | 73 | 1.06 | 2.4 |
| PSR J0248+6021 | 4035 | 58 | 0.96 | 2.1 |
| PSR J0357+3205 | 563 | 76 | 0.55 | 0.1 |
| PSR J0631+1036 | 832 | 66 | 0.51 | 1.7 |
| PSR J0633+0632 | 9720 | 61 | 1.07 | 1.2 |
| PSR J0659+1414 | 585 | 66 | 0.5 | 0.4 |
| PSR J1907+0602 | 3577 | 60 | 0.77 | 28.2 |
| PSR J1932+1916 | 1779 | 69 | 0.66 | 4.1 |
| PSR J1939+2134 | 1101 | 74 | 0.68 | 10.9 |
| PSR J1954+2836 | 457 | 75 | 0.9 | 10.5 |
| PSR J2021+3651 | 5393 | 72 | 0.81 | 33.8 |
| PSR J2021+4026 | 2069 | 67 | 0.78 | 1.1 |
| PSR J2032+4127 | 3842 | 70 | 0.85 | 2.7 |
| PSR J2229+6114 | 3884 | 57 | 0.65 | 223.1 |
| PSR J2238+5903 | 480 | 69 | 0.51 | 8.9 |
| PSR J2240+5832 | 340 | 69 | 0.8 | 2.2 |

**Table 1**: Summary of VERITAS observations. From left to right: source name, preliminary exposure time in minutes, average elevation in degrees, and preliminary average offset in degrees.

## 3. VERITAS Observations and Analysis

The Very Energetic Radiation Imaging Telescope Array System (VERITAS), located at the Fred Lawrence Whipple Observatory (FLWO) in southern Arizona (31 40N, 110 57W, 1.3km a.s.l.) is an array of four 12-meter diameter imaging atmospheric Cherenkov telescopes [17]. VERITAS is sensitive in the energy range of 85 GeV to >30 TeV. VERITAS has an energy resolution of 15%-25% at 1 TeV and a typical angular resolution of $< 0.1°$.

As mentioned in the previous section, the locations of 19 known pulsars have been observed by VERITAS with 11 locations having 20 hours or more of observations. After eliminating data affected by technical problems or taken under variable or poor sky conditions the total archival data set of known pulsars comprises more than 700 hours of observations carried out under four-telescope operations. A list of the observed targets and details of observations taken for each pulsar can be found in Table 1.

### 3.1 Phase-Gate Selection Procedure

The procedure for defining phase gates for pulsed emission searches in VERITAS data determines optimal phase gates to use for VHE pulsed analysis given an exposure time, an assumed VHE source flux and a phaseogram published in the supplementary material of the 2nd Fermi-LAT Pulsar Catalog (2PC) [14]. Event rates for signal and background regions are obtained from a reflected region analysis [18] of the Crab Nebula and event selection criteria that are optimal for sources with 1% Crab Nebula flux. Specifically, the rates of ON events ($N_{ON}$) and α-scaled OFF ($αN_{OFF}$) events are obtained. The rates are then multiplied by the VERITAS exposure time on a given pulsar to obtain the $N_{ON}$ and $αN_{OFF}$ events for the given exposure time. The number of excess events ($N_{excess}$) is found by $N_{excess} = N_{ON} - αN_{OFF}$. $N_{excess}$ is then scaled to 1% of the calculated value, corresponding to an assumed VHE source flux of 1% of the Crab nebula. The *Fermi*-LAT 2PC phaseogram for the source is used, and the value of the least populated bin is subtracted from all of the bins in the phaseogram in order to subtract the un-pulsed background. The background subtracted phaseogram is then normalized, and each bin of the background subtracted normalized phaseogram is multiplied by the scaled excess. Background counts determined from the offline analysis of the Crab Nebula are then added. Specifically, the number of $α$-scaled OFF counts divided by the number of bins in the phaseogram ($αN_{OFF}/N_{bins}$) is added to each bin of the phaseogram. All non-overlapping phase-gate combinations are tested with optional constraints on minimum and maximum phases and gate widths for each phase gate and number of signal phase gates. The number of events in the signal and background regions for a tested combination of gates is used to calculate a significance using equation (17) from [19]. The combination of gates that results in the highest significance is saved and plotted as shown in Figure 2. These phase gates will be used in the analysis of VERITAS archival data.

The number of signal phase gates for a given pulsar is chosen based on a visual inspection of the Fermi-LAT phaseogram. For 100 bin phaseograms, the maximum tested gate widths were constrained to be larger than the widest visually identified signal and background regions but smaller than the width of the entire phaseogram. For phaseograms with two peaks with different heights, phase gates are not defined for the small peaks using the procedure described above. A three pass procedure is used to define phase gates for the small peaks in these cases. The first pass consists of the method described above. The second pass defines only one signal phase gate, and excludes the gate defined for the large peak in the first pass from the signal gate search region. This ensures a gate is defined for the small peak. The third pass defines two signal phase gates but constrains one of the signal gates to be the one defined in the second pass. The three pass procedure effectively defines phase gates for the small peaks in these cases and was used to define the gates for PSRJ0205+6449 and PSRJ2021+4026.

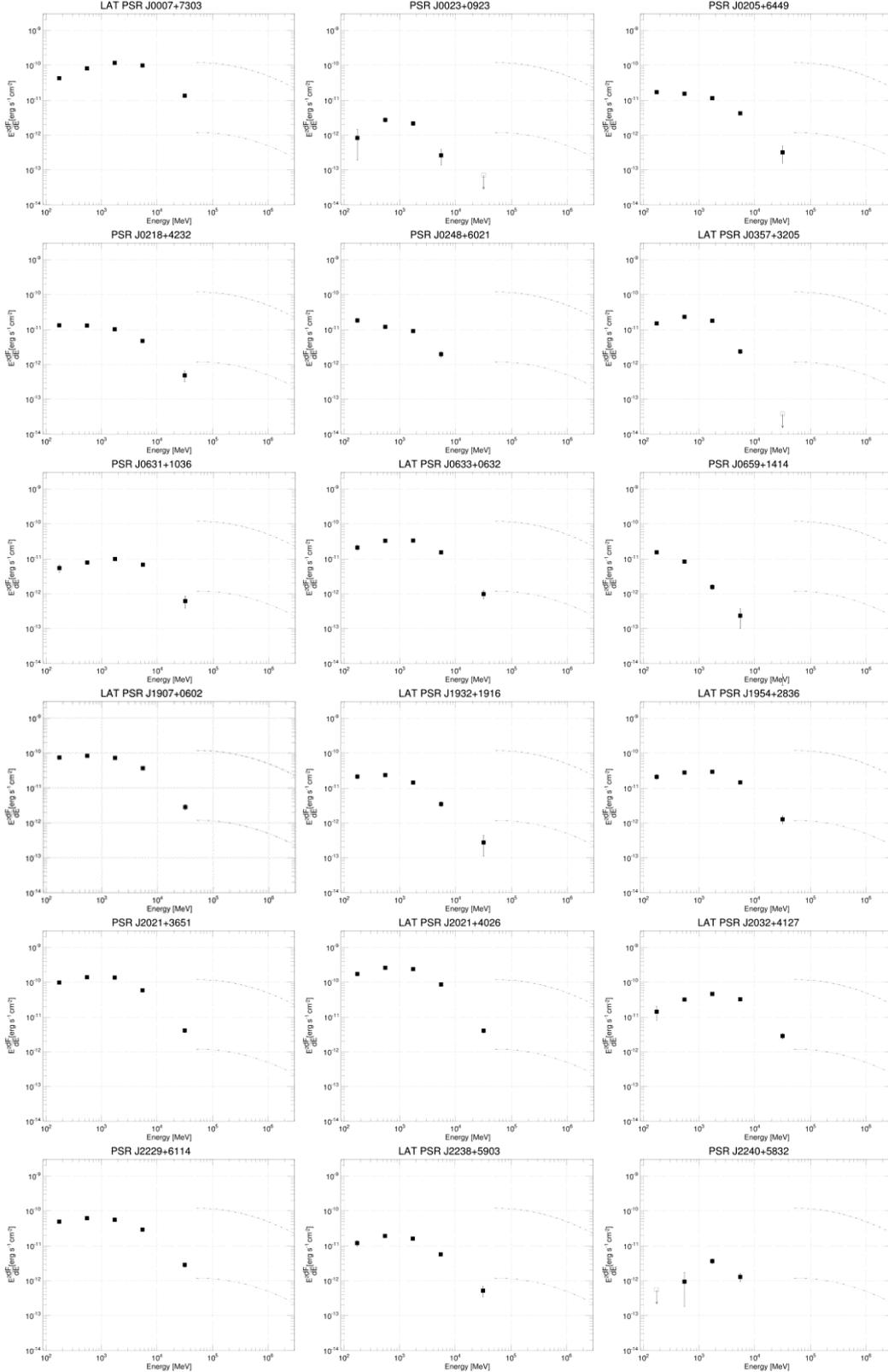

**Figure 1:** Phase-averaged SEDs from 3FGL for each of the 18 pulsars in VERITAS archival data (excluding PSRJ1939+2134). Measurements of 100% Crab Nebula flux and 1% Crab Nebula flux are plotted for comparison.

**4. Discussion and Future Plans**

Pulsar studies with VERITAS are an active and ongoing effort. Last year the collaboration developed a *Long Term Plan* to map out scientific priorities for the next five years and pulsar studies are one priority in the *Long Term Plan*. The VERITAS pulsed analysis of the pulsars in the VERITAS archival data is not fully completed and results of these analyses are forthcoming.

In addition to analysis of archival data, different populations of pulsars are being examined for targeted observations. One class of pulsar that may be of particular interest are millisecond pulsars. With very short spin periods, the light cylinders of these pulsars have smaller radii and thus more compact magnetospheres than other pulsars. Observations of millisecond pulsars can serve as another avenue to constrain VHE emission models and mechanisms. For more details on the constraints that millisecond pulsars can provide see the VERITAS contriubtion to this conference [20].

**5. Acknowledgements**

This research is supported by grants from the U.S. Department of Energy Office of Science, the U.S. National Science Foundation and the Smithsonian Institution, and by NSERC in Canada. We acknowledge the excellent work of the technical support staff at the Fred Lawrence Whipple Observatory and at the collaborating institutions in the construction and operation of the instrument. The VERITAS Collaboration is grateful to Trevor Weekes for his seminal contributions and leadership in the field of VHE gamma-ray astrophysics, which made this study possible.

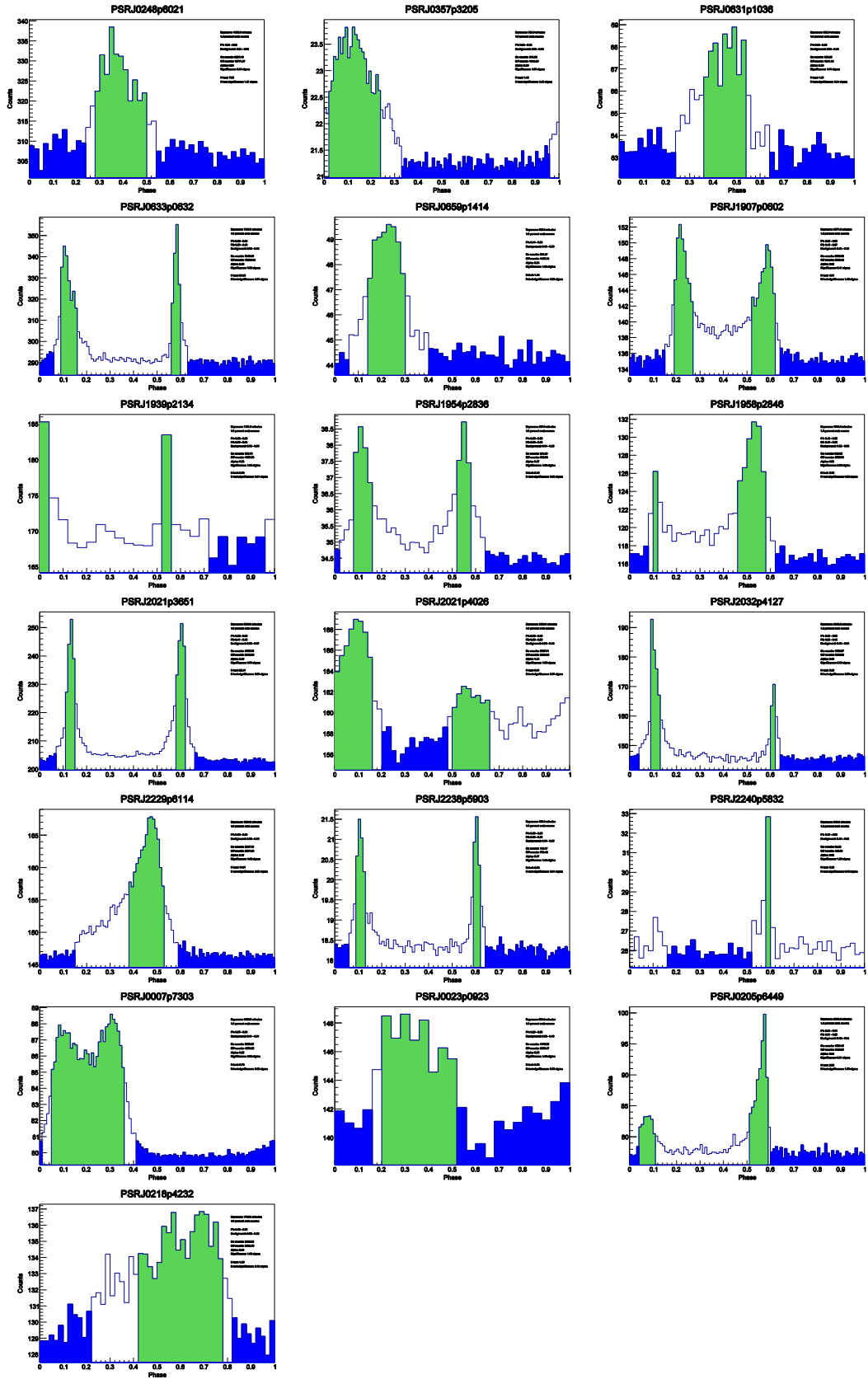

**Figure 2.** Simulated phaseograms of VERITAS archival pulsars with phase gates defined by the method described. P1 and P2 signal gates are shown in green, background gates are shown in blue.